\documentclass[fleqn,usenatbib]{mnras}

\usepackage{newtxtext,newtxmath}

\usepackage[T1]{fontenc}

\DeclareRobustCommand{\VAN}[3]{#2}
\let\VANthebibliography\thebibliography
\def\thebibliography{\DeclareRobustCommand{\VAN}[3]{##3}\VANthebibliography}

\usepackage{graphicx}	
\usepackage{amsmath}	
\usepackage{soul}

\newcommand{\angstrom}{\mbox{\normalfont\AA}}
\newcommand{\hi}{H{\sc i}}

\title[\rm HI at z=1.3]{Detection of \hi\ 21 cm  emission from a strongly lensed galaxy at $z\sim1.3$}

\author[Arnab Chakraborty]{
Arnab Chakraborty $^{1}$\thanks{E-mail: arnab.chakraborty2@mail.mcgill.ca, arnab.phy.personal@gmail.com}
and Nirupam Roy $^{2}$\\
$^{1}$Department of Physics and McGill Space Institute, McGill University, Montreal, QC, Canada H3A 2T8 \\
$^{2}$Department of Physics, Indian Institute of Science, Bangalore 560012,India 
}

\date{Accepted XXX. Received YYY; in original form ZZZ}

\pubyear{2015}

\begin{document}
\label{firstpage}
\pagerange{\pageref{firstpage}--\pageref{lastpage}}
\maketitle

\begin{abstract}
We report the first  $5\sigma$ detection of  \hi\ 21 cm  emission from a star-forming galaxy at redshift $z\sim1.3$  (nearly 9 billion years ago) using upgraded Giant Metrewave Radio Telescope (uGMRT). This is  the highest redshift  \hi\ detection  in emission from an individual galaxy to date. The emission is strongly boosted by the gravitational lens, an early type elliptical galaxy, at redshift $z \sim 0.13$.  The measured \hi\ mass of the galaxy is $\rm M_{HI} = (0.90 \pm 0.14 \pm 0.05) \times 10^{10}M_{\odot}$, which is almost twice the inferred stellar mass of the galaxy, indicating an extended structure of the \hi\ gas inside the galaxy.  By fitting two-dimensional Gaussian to the \hi\ signal at the peak of the spectral line, we find the source to be marginally resolved with the position angle consistent with the emission being tangential to the critical curve of the lens mass distribution.  This indicates that the solid angle of the approaching \hi\ line flux comes very close to  the inner lens caustic  and results in very high magnification. These results, for the first time, demonstrate the feasibility of observing high redshift \hi\ in a lensed system with a modest amount of telescope time and open up exciting new possibilities for probing the cosmic evolution of neutral gas with existing and upcoming low-frequency radio telescopes in the near future. 
\end{abstract}




\section{Introduction}

The reservoir of cold atomic neutral hydrogen (\hi) gas provides the basic fuel for star formation in a galaxy. Understanding the evolution of galaxies over cosmic time requires knowledge of the cosmic evolution of this neutral gas.  A detailed study of star formation history, over the last decade, shows that the comoving  star formation rate (SFR) density rises from $z\sim8$ to $z\sim3-4$, shows a peak, and remains flat in $z\sim3-1$, and then declines by an order of magnitude, from $z\sim1$ to the present epoch (e.g. \citealt{ LeFloch2005ApJ...632..169L,Hopkins2006ApJ...651..142H,Bouwens2009ApJ...705..936B,Madau2014ARA&A..52..415M}). Also the nature of galaxies undergoing star formation evolves significantly from $z\sim2$ to the present epoch \citep{Cowie1996AJ....112..839C}. At the peak of star formation ($z\sim 1-3$), the SFR density is dominated by massive galaxies with high SFRs, whereas in the local universe ($z\sim0$) it mostly arises in low mass systems with low SFRs \citep{LeFloch2005ApJ...632..169L}. 
 However the neutral H{\sc i} mass density ($\Omega_{\mathrm{H{\sc I}}}$) does not show any significant evolution over cosmic time \citep{Aditya2020Natur.586..369C,CHIME2022arXiv220201242C}. The molecular hydrogen ($\mathrm{H_{2}}$) density  also similar to SFR density shows a peak around $z \sim 1.5$ and then declines by one order of magnitude to present day \citep{Walter2020ApJ...902..111W}. On a contrary, the stellar mass density is increasing continuously with cosmic time and surpasses
the total gas density (\hi\ and $\mathrm{H_{2}}$ ) at redshift $z \sim 1.5$ \citep{Walter2020ApJ...902..111W}. This opposite nature of gas density and stellar mass density is puzzling and likely to be explained by the infall of ionized gas from IGM/CGM to the \hi\ reservoir and subsequent conversion of \hi\ to $\mathrm{H_{2}}$ \citep{Walter2020ApJ...902..111W}. \citet{Aditya2020Natur.586..369C} shows that accretion of  gas onto galaxies at $z \leq 1$ may have been insufficient to sustain high star-formation
rates in star-forming galaxies and likely to be the cause of the decline in the cosmic
star-formation rate density at redshifts below one. However, the evolution of cold neutral gas during $z \sim 0-3$ still needs to be constrained with more sensitive observation to understand the global flow of gas onto galaxies and to probe the history of star formation in the Universe. Hence  the knowledge of \hi\ mass of different types of galaxies and the relation between the atomic and molecular gas mass, stellar mass, and the star formation rate, is critical to study galaxy evolution.

The best way to probe the neutral atomic gas content in a galaxy is via the \hi\ 21 cm spectral line emission. However,  the probability for spontaneous emission of \hi\ 21 cm radiation, due to the spin-flip transition between hyperfine states in the ground state of neutral hydrogen, is extremely low. Due to this, it is  challenging to detect the \hi\ line emission from  individual galaxies beyond a redshift of about 0.4,  with modern telescopes.  The highest redshift detection of \hi\ line emission to date from an individual galaxy was made  at $z = 0.376$ \citep{Fernandez2016ApJ...824L...1F}. The average properties of \hi\ content of galaxies can be obtained by `stacking' the \hi\ line emission signals of a large number of galaxies with known spectroscopic redshifts \citep{Chengalur2001A&A...372..768C,Zwaan2000PhDT..........Z}. This method has been used to measure the average \hi\ mass and cosmological \hi\ mass density at higher redshifts, $z\gtrsim1$ \citep{Aditya2020Natur.586..369C,Aditya2021ApJ...913L..24C}, but it is not possible to measure the properties of individual sources via `stacking'. The measurement of \hi\ masses of individual galaxies at $z\gtrsim1$ would require a long integration time with today's radio telescope or the large collecting area of the Square Kilometer Array (SKA).

The strong gravitational lens, nature's gift, magnifies the weak emission signal coming from distant objects, enabling us to peer through the high redshift universe. Strong gravitational lensing phenomenon can significantly amplify the faint signal, enabling us to detect the \hi\ signal from galaxies at higher redshifts  in a reasonable observation time.   At small wavelengths (mm-wave), the amplification of the faint signal from a distant galaxy through gravitational lensing has been used to observe the high redshift universe (e.g. \citealt{Brown1991AJ....102.1956B,Vieira2013Natur.495..344V}).   \citet{Hunt2016AJ....152...30H} made the first attempt to detect the \hi\ 21 cm signal from a lensed galaxy at $z\sim0.4$. They tried to detect the \hi\ signal from three  
lensed galaxies, two at $z=0.398$ and one at $z=0.487$, using the Green Bank Telescope (GBT). The background galaxies are lensed by the foreground cluster Abell 773. However, they did not detect the signal and have reported a $3\sigma$ upper limit on the \hi\ mass of the galaxies \citep{Hunt2016AJ....152...30H}. \citet{Blecher2019MNRAS.484.3681B} have also tried to detect the \hi\ signal from a galaxy at $z\sim 0.4$, using gravitational lensing. They also did not detect the signal with a high signal-to-noise ratio. They estimate the \hi\ mass, using Bayesian formalism, from the integrated \hi\ spectrum. There is no strongly lensed \hi\ detection in emission with high statistical significance to date \citep{Blecher2019MNRAS.484.3681B,Hunt2016AJ....152...30H}.

In this paper, we report the first detection of \hi\ 21 cm line emission from a galaxy, a galaxy-galaxy strong lens candidate at $z\sim 1.3$, which is detected in Sloan Lens ACS (SLACS) Survey for the Masses (S4TM) Survey \citep{Shu2017ApJ...851...48S},  using GMRT archival data. This is  the highest redshift  \hi\ detection  (lookback time $\sim$ 9 Gyr) from an individual galaxy to date.  The paper is organized as follows: we mention the details of the target galaxy in Sec. \ref{details_of_the_target}, we describe our data analysis and estimated \hi\ spectrum in Sec. \ref{sec:obs}, the estimation of \hi\ mass, atomic-to-stellar mass ratio and the extension of the \hi\ emission are mentioned in Sec. \ref{sec: HI mass}, finally, we conclude in Sec. \ref{sec:conclusion}. Throughout this work, we use  the Planck 2015 cosmological parameters \citep{Planck2016A&A...594A..13P}.

\begin{table}
\centering
\begin{tabular}{ll}
\hline
Source redshift ($z_{S}$) & 1.2907 \citep{Shu2017ApJ...851...48S}\\

Lens redshift ($z_L$) & 0.1318 \citep{Shu2017ApJ...851...48S} \\

Optical magnification & 105 \citep{Shu2017ApJ...851...48S} \\

Einstein radius & $1.01''$ \citep{Shu2017ApJ...851...48S} \\

Position angle of lens & $82^{\circ}$ \citep{Shu2017ApJ...851...48S} \\

\hi\ magnification ($\mu_{\rm HI}$) & $29.37 \pm 6$ \\

\hi\ mass ($M_{\rm HI}$) & $(0.90 \pm 0.14 \pm 0.05) \times 10^{10}M_{\odot} $ \\

 Inferred stellar mass  ($M_{*}$) & $(0.38 \pm 0.11) \times 10^{10}M_{\odot}$  \\

$M_{\rm HI}/M_{*}$ & $2.37 \pm 0.14$ \\
\hline
\end{tabular}
\caption{\label{tab:example} The details of the  target galaxy and key findings. }
\end{table}

\section{Details of the target galaxy}
\label{details_of_the_target}
 The target source was selected from the catalog of the galaxy-galaxy strong-lens candidates, detected in the Sloan Lens ACS (SLACS) Survey for
the Masses (S4TM) Survey \citep{Shu2017ApJ...851...48S}.   The S4TM survey was designed to identify low to intermediate-mass  Early-type galaxies (ETG), which act as a strong lens system. The S4TM survey detected 118 nearly  strong lens candidates, selected spectroscopically from the galaxy spectrum database of the seventh and final data release of the Sloan Digital Sky Survey (SDSS).  The basic method to select such lens candidates is to search for multiple nebular emission lines in the spectrum coming from a common redshift, which is significantly higher than the redshift of the foreground  lensing candidate.   This indicates that there are two objects within the same lightcone of diameter 3 arcsec (diameter of the optical fiber) and a lensing event happened \citep{Bolton2004AJ....127.1860B}.  The candidates were further observed with Hubble Space Telescope (HST) in the F814W-band.  \citet{Shu2017ApJ...851...48S} modeled the  foreground light of the lens galaxy with an elliptical radial B-spline model and subtracted it from the image. The residual image was then inspected for lens morphology, multiplicity, and lens grade. Any candidate was classified as a grade-A strong lens, if definite multiple lensed images were detected \citep{Shu2015ApJ...803...71S}. There were 40 such grade-A strong lens candidates discovered for the first time \citep{Shu2017ApJ...851...48S}. These grade-A lens candidates were modeled as singular isothermal ellipsoid (SIE) profile \citep{Kormann1994A&A...284..285K} and the background source light distribution was modeled as multiple elliptical Sersic components.   Using the magnification factor, redshift, and the luminosity distance, we found out the value of the quantity $\big(\frac{D_{L}^{2}}{\mu_{\rm HI}(1+z)}\big)$  for each of the candidate source. The velocity integrated \hi\ flux is inversely proportional to this quantity for a fixed \hi\ mass (see Eqn. \ref{mass_HI}). Hence, the \hi\ signal from a source galaxy is  more likely to be detected if this quantity is less for that source. Then we ranked these candidates based on this value, i.e, a source is more likely to be detected with a moderate telescope time if the value of this quantity is smaller for a fixed  \hi\ mass.
 Then we searched the archival data of GMRT for the  first five sources from this list and found that the top-ranked candidate was observed with uGMRT in cycle-34 (Proposal code- 34\_066).  For this  target system, the lensing galaxy (SDSSJ0826+5630) shows an average optical magnification of about 105 and is situated at $z_{L} = 0.1318$, whereas the background source galaxy was at $z_{s} = 1.2907$. This galaxy-galaxy lens system shows a nearly full Einstein ring with a radius of about $1.02''$ \citep{Shu2017ApJ...851...48S}. This is a very promising source because - (i) the target source galaxy shows multiple nebular line emissions, which signifies that it is a star-forming galaxy. Also, the redshift of the source galaxy ($z_{S}$ $\sim$ 1.3) is close to the peak of the star-formation rate density. Hence, the galaxy is expected to have a significant cold neutral gas reservoir, (ii) an extremely strong optical magnification suggests that the \hi\ magnification will also be large and there will be a strong boost of the \hi\ flux coming from this distant background source galaxy.  Hence,   it may be possible to  detect the \hi\ 21 cm emission from the background galaxy with moderate observing time, although the source galaxy  resides at a large cosmological distance from us, (iii) the mass distribution of the lens is already modeled precisely using deep HST optical data set. Hence, we can use those model parameters for the analysis of the \hi\ magnification. Details of different model parameters are mentioned in Table \ref{tab:example}.
 
\section{Observation, data analysis and results} \label{sec:obs}

 The  galaxy, with pointing center at  $\rm RA =08^{h}26^{m}39.858^{s},\rm DEC =56^{\circ}30'35.97''$,  was observed with uGMRT Band-4 receivers for a total of 18 hours on-source time.  A bandwidth of 100 MHz, sub-divided into 2048 channels, was used for the observation with GMRT wideband backend (GWB) as the correlator. The frequency coverage was 550-650 MHz, with a velocity resolution of $\sim$24 km/s.  The integration time per visibility point was 5s.
The standard calibrators, 3C147 and 3C286, were observed to calibrate the flux density scale, while regular observations of the nearby compact source 0834+555  were used to calibrate the complex antenna gains.

The data was first inspected using AOFLAGGER \footnote{\url{https://aoflagger.readthedocs.io/en/latest/}} package for the detection and excision of radio frequency interference  \citep{Offringa2012A&A...539A..95O}. We used a {\sc casa}  based flagging and calibration pipeline  to  solve the complex gains  and to remove any remaining bad data, following the standard procedure.  We used the automatic algorithms, {\sc tfcrop} and {\sc rflag}, within {\sc casa}'s {\sc flagdata} task  to identify and remove the radio frequency interference (RFI).  The flux density of the primary calibrators, 3C147 and 3C286, were set using the Perley-Butler model \citep{Perley2017ApJS..230....7P}. The delay and bandpass corrections were derived from the observations of the primary calibrators. The time variable complex gains for each antenna were derived from the observation of secondary calibrator 0834+555,  bf which was frequently observed for 2 mins for every 15 mins scan of the target. Following this, the calibration solutions were applied to the target field and we split the target for imaging and self-calibration.  We did not average the data across frequency and time and retain the maximum resolution. This helped us to identify and flag the  bad data during self-calibration and imaging loops.

\begin{figure}
\centering

   \includegraphics[width=\columnwidth]{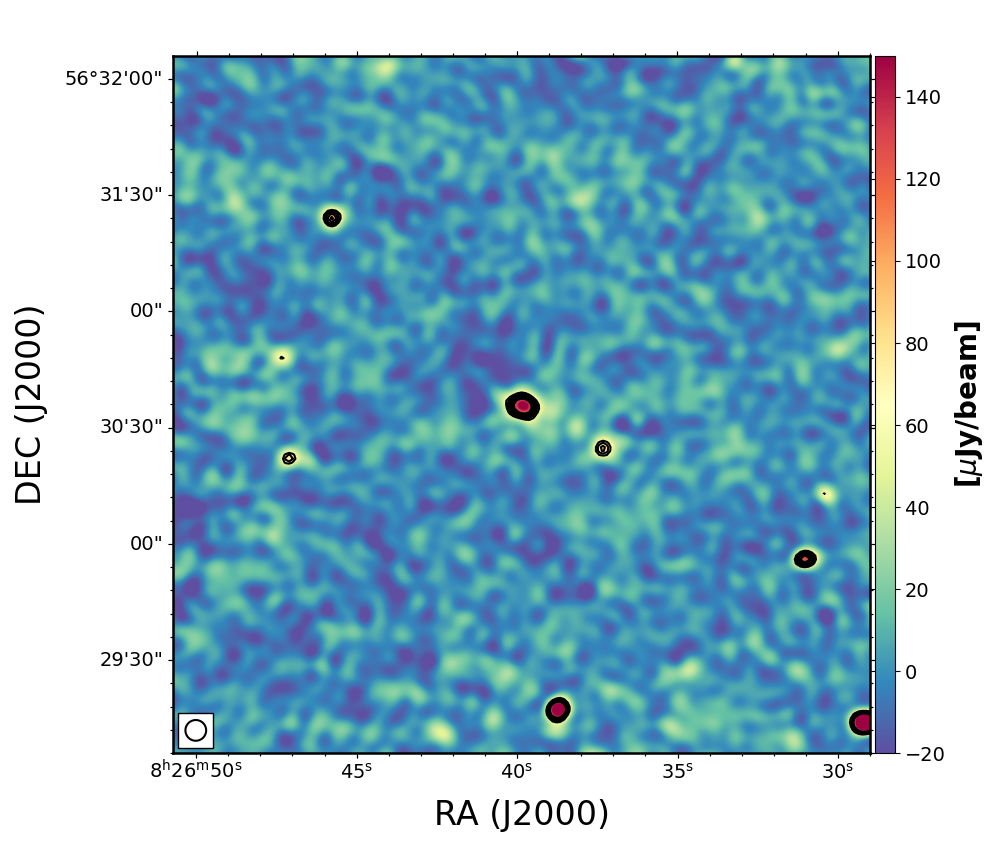}  

\caption{The continuum image of the target field at 600 MHz,  showing the  central $3' \times 3'$ area. The ellipse at the bottom left is the synthesized beam, with major and minor axes $4.5'' \times 4.4''$, and the position angle is $82^{\circ}$. The contour levels are at -4, 5, 6, 7, 8, 9, and 10$\sigma$ (negative contour is in dashed line) statistical significance, where $\sigma$ = 8 $\mu$Jy beam$^{-1}$ is the RMS noise on the continuum image near the phase center.  We do not find anything at the $-4 \sigma$ level in this region.}
\label{fig:cont}
\end{figure}

 We used {\sc wsclean} \citep{Offringa2014MNRAS.444..606O} to make the continuum image of the target field.  The multi-scale wide-band deconvolution along with auto-masking \citep{Offringa2017MNRAS.471..301O} was performed  to capture the variation of sky brightness across this large bandwidth over different spatial scales.  We made a large image of size  8192$\times$8192 pixels, covering a total field of view of $2.27^{\circ} \times 2.27^{\circ}$, with a pixel size of $1.0''$.  This large image is required to  deconvolve and model the bright confusing sources far away from the first null of the primary beam.  We  made the first image down to 6$\sigma$ using the auto masking routine of {\sc wsclean}. The deconvolution was terminated after 50k iterations. Then we created a mask using the first image  down to 10$\sigma$  to remove any spurious features. Then we run another  constrained deconvolution using that mask file in order to generate an artifact-free model for self-calibration purposes.

{\sc wsclean} inverts the frequency-dependent skymodel derived from the deconvolution process into model visibilities at the end of the imaging process, which we used for the self-calibration. We performed several rounds of phase-only self-calibration, with an improved mask at each iteration, until no further improvements were seen in the continuum image. We used Briggs weighting with a robust parameter of -1 during self-calibration and the final continuum image was made using the robust parameter of 0.0 \citep{Briggs1995AAS...18711202B}. The  continuum image was shown in Fig.\ref{fig:cont}. The off-source RMS noise near the field center, away from the bright source,  is about 8$\mu$Jy beam$^{-1}$ with  a synthesized beam width of about $4.5''$.

\begin{figure}
\centering

   \includegraphics[width=\columnwidth]{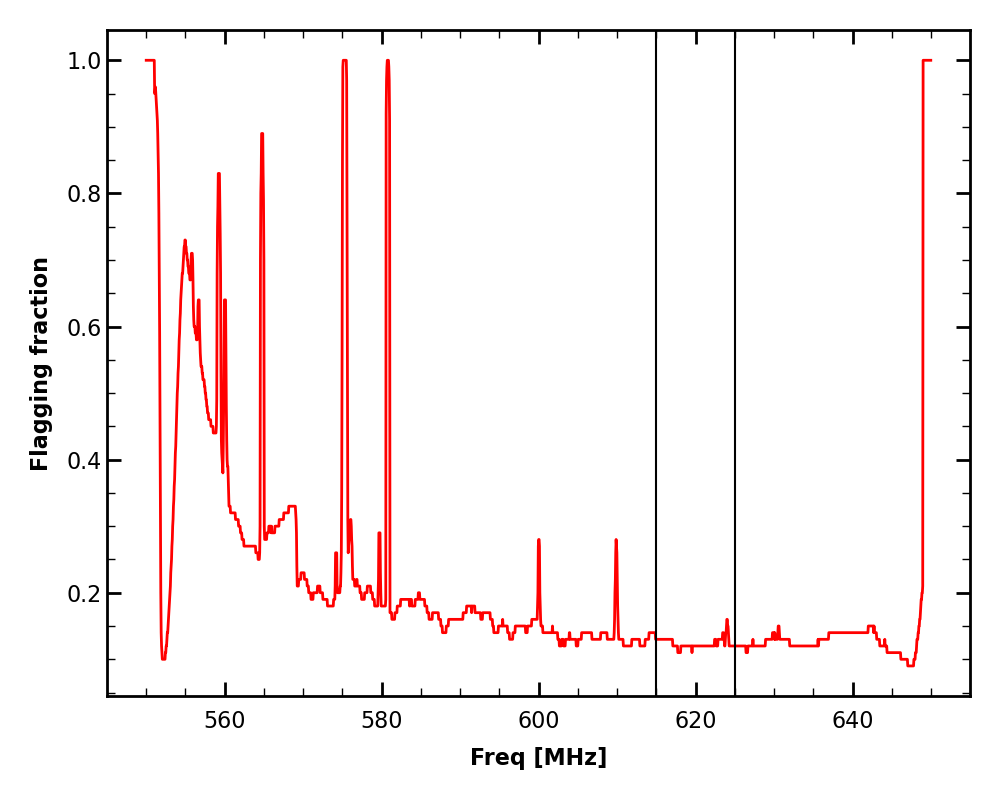}  

\caption{Fraction of flagging or lost data due to RFI mitigation as a function of frequency. The region bounded by the vertical black lines  is being used for line-cube analysis.}
\label{fig:flag_frac}
\end{figure}

We then subtracted the continuum emission from the calibrated multi-channel visibilities using {\sc uvsub} routine in {\sc casa}. Then, any residual continuum emission was subtracted via a 2-nd order polynomial fit to each visibility spectrum  and the residual visibilities were then shifted to barycentric frame, using the {\sc mstransform} routine in {\sc casa}.  We excluded  50 channels on each side of the central line frequency channel  $\nu = 620.0683$ MHz, corresponding to $z=1.2907$, during the polynomial fitting to the visibility spectrum.  
 We have also checked the data visually in the time-frequency plane for any residual RFI close to the line-frequency,  using {\sc rfigui} routine of AOFLAGGER,  but did not find any such contaminants.  The fraction of data lost due to RFI mitigation as a function of frequency is shown in Fig. \ref{fig:flag_frac}. The vertical black lines show the region, which is being used for the final line-cube analysis.  We made a spectral image cube using {\sc tclean} routine in {\sc casa}, with natural weighting and w-projection algorithm \citep{Cornwell2008ISTSP...2..647C}. We used the baselines $<18 \rm K \lambda$ and a Gaussian uv-taper at $12 \rm K \lambda$ during imaging. This gave us the optimal spatial resolution of about $13''$ to extract the \hi\ signal with the highest signal-to-noise ratio. This spatial resolution corresponds to the physical size of about 112 kpc at the redshift of the background galaxy. The resulting spectral image cube has a frequency resolution of  48.83 kHz, corresponding to a velocity resolution of 24 km/s.

We take a cut along the velocity axis at the central peak position of this spectral cube and the resultant spectrum is shown in the left panel of Fig. \ref{fig:HI_Spectrum}.  The black line is the $1\sigma$ uncertainty on the spectrum. We performed three different tests to estimate the $1\sigma$ RMS noise on the spectrum. First, we take the spectrum corresponding to the line free channels, i.e, by excluding the three channels around the central peak channel as seen in Fig. \ref{fig:HI_Spectrum}.  Then we perform the Anderson-Darling test on this line-free spectrum to check for Gaussianity. The null hypothesis is that the line-free spectrum corresponds to the Gaussian distribution. The estimated $p$ value is 0.16, which shows that the line-free spectrum corresponds to the Gaussian noise distribution and we quote the   RMS of this line-free spectrum  as $1\sigma$ uncertainty here. The $1\sigma$ RMS  noise is $\sim 154$ $\mu$Jy beam$^{-1}$, which is being shown by black line in Fig. \ref{fig:HI_Spectrum}. Next, we take an off-source region of size $8$ times the synthesized beam close to the phase center and estimate the RMS for this region along the frequency axis. We found that the estimated RMS is consistent with our previous finding. In addition to this, we also take spectrum along 50 arbitrarily chosen line-of-sight through the spectral cube and found that the mean of those spectra is also consistent with our estimation of $1\sigma$ uncertainty on the spectrum. 
{ It is clear from the left panel of  Fig. \ref{fig:HI_Spectrum} that the \hi\ 21 cm  peak flux corresponding to the central channel ($\nu = 620.0683$ MHz)  from the background galaxy ($z_{s}\sim1.3$) is detected at  4$\sigma$ significance.
The average map of the three channels, the central peak channel, and the two neighboring channels,  is shown in the right panel of Fig. \ref{fig:HI_Spectrum}. The contours are at $[-4,4,4.25,4.5,5] \times \sigma$ levels, where $\sigma$ = 108 $\mu$Jy beam$^{-1}$ is the RMS noise of the channel average image. The synthesized beam of the channel average image, with major and minor axes $13.37''\times12.32''$ and position angle $37^{\circ}$,  is shown in the bottom left. 
 The channel average  image also shows a clear detection of the \hi\ 21 cm emission signal from the distant background galaxy at $5\sigma$ significance.

\begin{figure*}

\begin{tabular}{cc}
  \includegraphics[width=3.25in,height=2.65in]{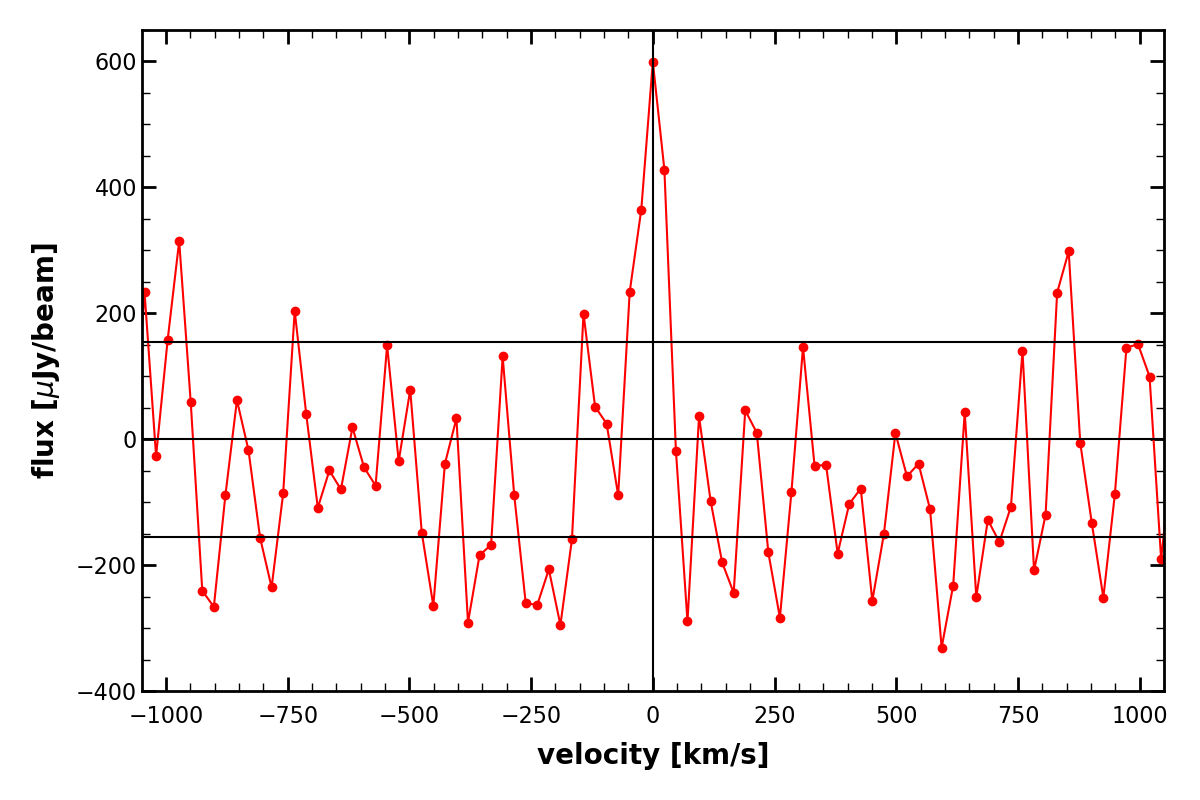} &
   \includegraphics[width=3.25in,height=2.68in]{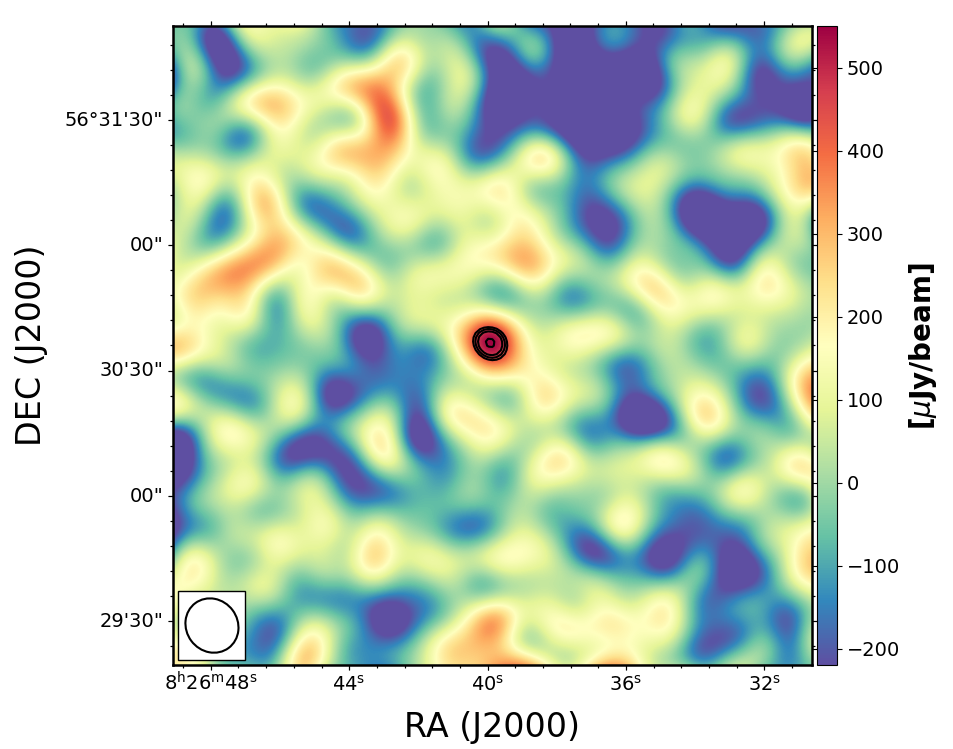}
  
\end{tabular}
\caption{Left: The \hi\ 21 cm emission spectrum   extracted along the peak pixel position of the galaxy  at a velocity resolution of 24 km $s^{-1}$. The black line shows the $1 \sigma$ uncertainty on the spectrum. Right:  The average image of the three channels (620.0194 - 620.1171 MHz) centered around the peak channel, $\nu = 620.0683$ MHz ($z = 1.2907$) of the spectral cube. The ellipse at the bottom left is the synthesized beam, with major and minor axes $13.37'' \times 12.32''$, and the position angle is $37^{\circ}$. The contour levels are at $[-4,4,4.25,4.5,5] \times \sigma$ statistical significance, where $\sigma$ = 108 $\mu$Jy beam$^{-1}$ is the RMS noise of the image. The image size is $2.55' \times 2.55'$, which corresponds to about $1315$ kpc at the redshift of the background galaxy.  The \hi\ emission signal, from the galaxy at $z_{s}\sim 1.2907$, is clearly detected in the center of the image, at $ 5\sigma$ significance.}
\label{fig:HI_Spectrum}
\end{figure*}

\section{Estimation of  {\hi\ Mass}} \label{sec: HI mass}

 The \hi\ magnification factor is not the same as the optical magnification as reported in \citet{Shu2017ApJ...851...48S}. The \hi\ mass and size are tightly correlated  and the mass increases linearly with the size of the \hi\ disk \citep{Wang2016MNRAS.460.2143W}. In general, the distribution of \hi\ in a galaxy is more extended than the stellar component, and as magnification is approximately equal to the lensed to the intrinsic angular size of a source, the \hi\ magnification is typically lower than the optical magnification \citep{Blecher2019MNRAS.484.3681B}. We performed simulations to estimate the \hi\ magnification  factor ($\mu_{\rm HI}$)   to infer the \hi\ mass from the measurement of lensed velocity integrated \hi\ flux. We followed the methodology presented in \cite{Blecher2019MNRAS.484.3681B}  for the simulation and only briefly mentioned it here.

 We first simulated an \hi\ disk, where the  intrinsic \hi\ surface density, $\Sigma_{\rm HI}$, was modeled as an axially symmetric surface density profile, given by  \citet{Obreschkow2009ApJ...698.1467O}, 

\begin{equation}
\Sigma_{\rm HI} (r) = \frac{M_{\rm  H}/(2\pi r_{\rm disk}^{2}) \exp{(-r/r_{\rm disk})}}{1+R^{\rm c}_{\rm mol}\exp{(-1.6 r/r_{\rm disk})}},
\label{eq:HI_profile}
\end{equation}

where $ M_{\rm  H}$ corresponds to the total hydrogen mass, i.e. atomic hydrogen mass ($M_{\rm HI}$)  plus the molecular hydrogen mass ($M_{\rm H_2}$);   $r$ denotes the galactocentric radius in the plane of the disk,   $r_{\rm disk}$ is the scale length of the neutral hydrogen disk and $R^{\rm c}_{\rm mol}$ corresponds to the ratio of molecular to atomic hydrogen mass given by \citet{Obreschkow2009ApJ...698.1467O}
\begin{equation}
 M_{\rm H_2}/M_{\rm HI} = (3.44 R^{\rm c\ -0.506}_{\rm mol}+4.82R^{\rm c\ -1.054}_{\rm mol})^{-1}.
\end{equation}

The \hi\ mass is strongly correlated with the \hi\ size given as  \citet{Wang2016MNRAS.460.2143W}, 

\begin{equation}
\log_{\rm 10}(D_{\rm HI})  = 0.506 \log_{\rm 10}(M_{\rm HI}) - 3.293,
\label{eq:mass_size}
\end{equation}
where $D_{\rm HI}$ is defined as the diameter at which the \hi\ density drops to $\Sigma_{\rm HI} = 1~{\rm M_\odot pc^{-2}}$. Note that here  $D_{\rm HI}$ is in units of kpc and $M_{\rm HI}$ is in units of ${\rm M_\odot pc^{-2}}$.

In our simulation, we sampled $\log_{\rm 10}(R^{\rm c}_{\rm mol})$ from a normal distribution with $[{\rm mean, stdev}] = [-0.1,0.3]$, which is consistent with the range of molecular to atomic gas mass ratio,  $M_{\rm H_2}/M_{\rm HI} \sim 0.12-0.32$,  for the stellar mass range $\rm log_{10} M_{*} \sim 9.18-11.20$ at $z=0$ \citep{Catinella2018MNRAS.476..875C}. The  \hi\ mass  is sampled between $\rm log_{10}M_{\rm HI} \sim 6-12$, which is consistent with the stellar mass range defined in \citet{Maddox2015MNRAS.447.1610M}. 
For a given $M_{\rm HI}$ and $R^{\rm c}_{\rm mol}$, we first find out  the value of $D_{\rm HI}$ using Eqn. \ref{eq:mass_size} and then use this to solve for $r_{\rm disk}$ in Eqn. \ref{eq:HI_profile}. To incorporate the orientation effects, the simulated \hi\ disk is rotated in a 3-dimensional cube to sample the position and inclination angle of the disk. The inclination angle ($i$) was sampled with probability density function (PDF)  of $sin(i)$ over the range [0,$\pi/2$], and the position was sampled randomly between [0,$\pi$].

\begin{figure}
\centering
  \includegraphics[width=\columnwidth]{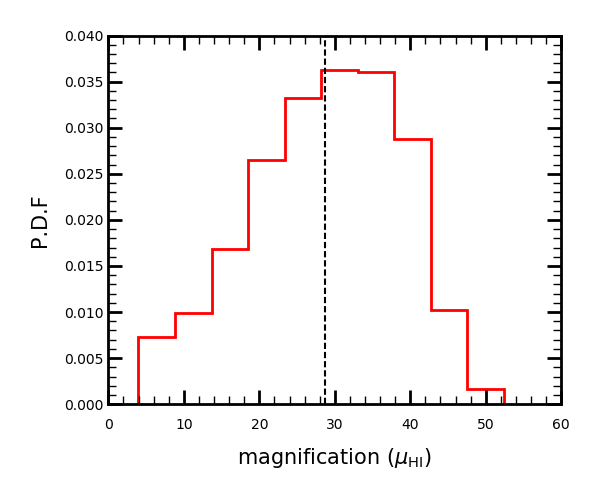}  

\caption{The probability distribution function of the \hi\ magnification factor ($\mu_{\rm HI}$) is shown. The black dashed line is the mean value of $\mu_{\rm HI}$. }
\label{fig:mag}
\end{figure}

We used the singular isothermal ellipsoid (SIE) profile to model the projected lens-mass distribution  as described in the S4TM survey \citep{Shu2017ApJ...851...48S}. The SIE model has two-dimensional mass density profile given as  \citep{Kormann1994A&A...284..285K},

\begin{equation}
\Sigma(x,y) = \Sigma_{\rm crit} \frac{\sqrt{q}}{2}\frac{b_{\rm SIE}}{\sqrt{x^{2}+q^{2}y^{2}}},
\label{eqn:sie}
\end{equation}

where $q$ is the minor-to-major axis ratio, $b_{\rm SIE}$ is the Einstein radius, $\Sigma_{\rm crit}$ is the critical density determined as 

\begin{equation}
    \Sigma_{\rm crit} = \frac{c^{2}}{4 \pi G} \frac{d_{S}}{d_{L}d_{LS}},
\end{equation}
where $d_{S}, d_{L}$ and $d_{LS}$ are the angular diameter distances from the
observer to the lens, from the observer to the source, and between
the lens and the source, respectively. The lens model does  not include any external shear as it is a minor effect  \citep{Shu2017ApJ...851...48S}. The Einstein radius, position angle, and ellipticity of the lens were set to that of the observed optical distribution \citep{Shu2017ApJ...851...48S}. 

\begin{figure*}
\centering

   \includegraphics[width=2.\columnwidth]{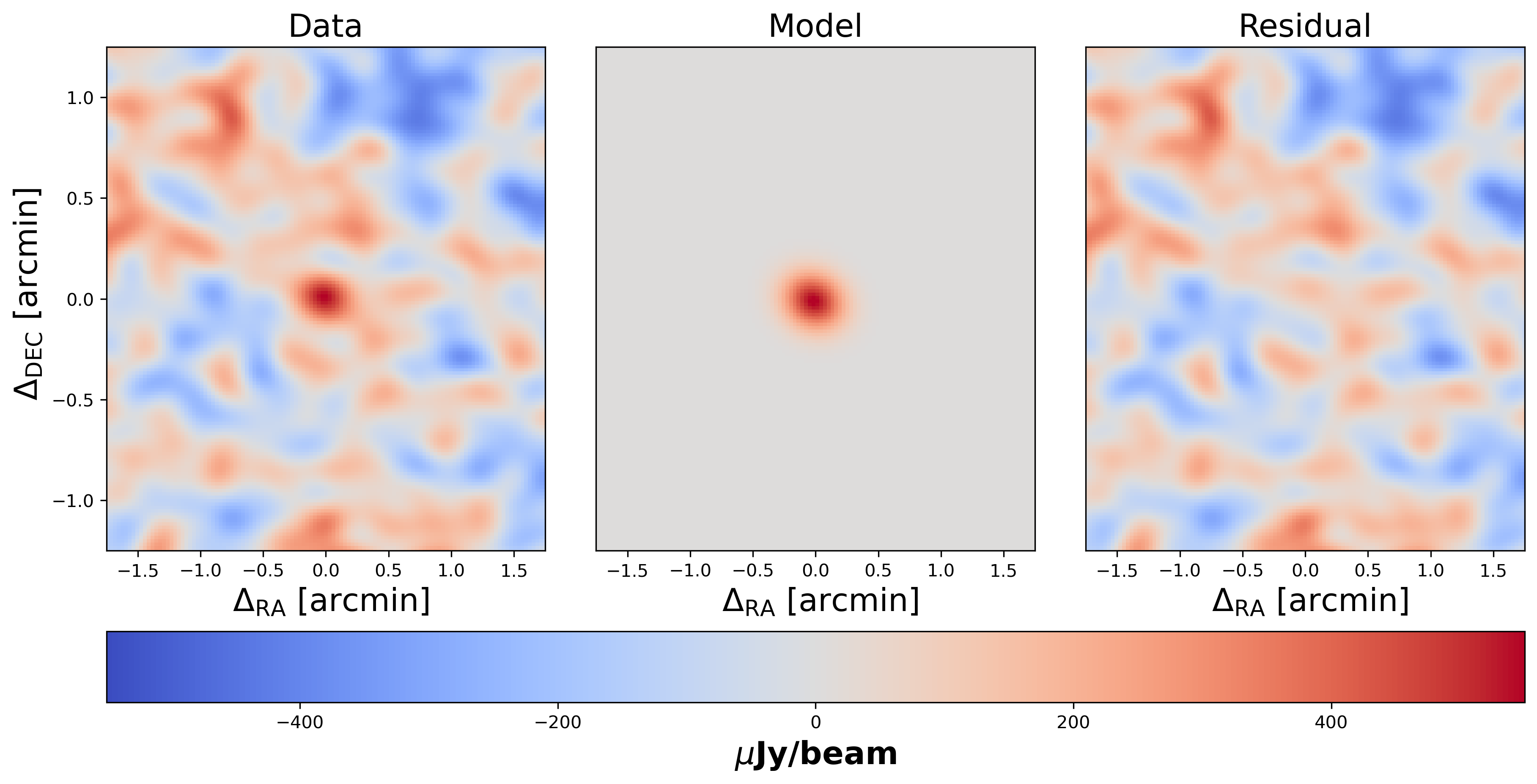}%

  \caption{In left we show the channel averaged image (data) same as the Fig. \ref{fig:HI_Spectrum}, the middle panel shows the simulated model of the source galaxy at $z\sim 1.3$ and the right panel shows the residual after subtracting the model from the image.}
\label{fiited_model}
\end{figure*}

The general relativistic ray tracing was performed using the GLAFIC package \citep{Oguri2010PASJ...62.1017O}. The position of the centroid of the source galaxy with respect to the lens, called the impact factor, was not known apriori  \citep{Shu2017ApJ...851...48S}. We varied the impact factors between [0.0-0.5] to yield the published optical magnification.

We ran $10^{4}$ Monte Carlo simulations, by varying the model parameters, and  estimate the magnification factor,  which yield the velocity integrated observed \hi\ flux.  We found that the  \hi\ magnification factor  entirely depends on the \hi\ mass and does not show any dependence on $R^{\rm c}_{\rm mol}$, inclination angle, and impact factor in our simulations.  The magnification is approximately equal to the ratio of  the lensed to the intrinsic angular size of the source.  As shown in Eqn. \ref{eq:mass_size},  the \hi\ mass is a  increasing function with the \hi\ size, hence the magnification strongly depends on the \hi\ mass. Note that, \citet{Blecher2019MNRAS.484.3681B} also found a similar behavior of magnification in their analysis. Hence, we marginalized over all other nuisance parameters and the 1-dimensional PDF of the \hi\ magnification factor  ($\mu_{\rm HI}$) is shown in Fig. \ref{fig:mag}.   The mean value of $\mu_{\rm HI}$ with $1\sigma$ error bar is $29.37 \pm 6$. In fig.\ref{fiited_model}, we show the data (left panel), the simulated model (middle panel), and the residual (right panel). We see that the simulated model captures the \hi\ emission of the source galaxy accurately. 

The \hi\ mass was estimated from the lensed spectrum, using the \hi\ magnification factor  as (see ref. \citealt{Wieringa1992A&A...256..331W} for  un-lensed galaxy), 

\begin{equation}
    \frac{M_{\rm HI}}{M_{\odot}} = \frac{1}{\mu_{\rm HI}} \frac{236}{(1+z)} \Big(\frac{D_{L}}{\rm Mpc}\Big)^{2} \Big(\frac{\int S_{V}dV}{\rm mJy km s^{-1}}\Big)
    \label{mass_HI},
\end{equation}
where $z$ is the redshift of the background source galaxy, $D_{L}$ is the luminosity distance in units of Mpc and $\int S_{V}dV$ is the velocity integrated \hi\ flux in units of $\rm mJy$ $\rm km$ $\rm s^{-1}$. 
To quote the uncertainty in the estimation of the mass, we first fit the line emission spectrum with a Gaussian and subtracted it from the entire spectrum.  The residual spectrum is consistent with the Gaussian noise. Then  we smoothed the residual spectrum to  50 km $s^{-1}$ velocity resolution (twice the original resolution). We take the $1\sigma$ of the smoothed residual spectrum and  use Eqn. \ref{mass_HI} to find out the mass within the same velocity window. The estimated mass by this process was quoted as uncertainty in the measured \hi\ mass of the galaxy. This method propagates the spectral RMS noise of the observed HI spectrum to the uncertainty in the HI mass and also ensures that any non-detection of \hi\ flux due to lower SNR inside the velocity window will be accounted for in the measurement uncertainty of \hi\ mass.   The  estimated \hi\ mass is $\rm M_{HI} = (0.90 \pm 0.14 \pm 0.05) \times 10^{10}M_{\odot}$, where the first uncertainty is due to the uncertainty in \hi\ magnification factor and the second one is due to the non-detection of the \hi\ flux because of low SNR as described above.

The most precise measurement of the \hi\ mass function (HIMF) to date using Arecibo Legacy Fast ALFA  (ALFALFA) catalog, at $z\sim0$, shows a power law-like increase towards the  lower masses and a sharp exponential decline  towards the higher mass end \citep{Jones2018MNRAS.477....2J}. This transition to the exponential decrease of the HIMF happened around a `knee' mass, $\rm M_{\rm knee} = 0.87 \times 10^{10}M_{\odot}$ \citep{Jones2018MNRAS.477....2J}. This was  found after fitting the Schechter  function to the measured HIMF of local galaxies detected in the ALFALFA survey. Hence, the estimated \hi\ mass of the lensed source galaxy  in this analysis falls close to the  `knee' mass of the  HIMF at $z=0$.

\begin{figure*}
\centering

   \includegraphics[width=2.\columnwidth]{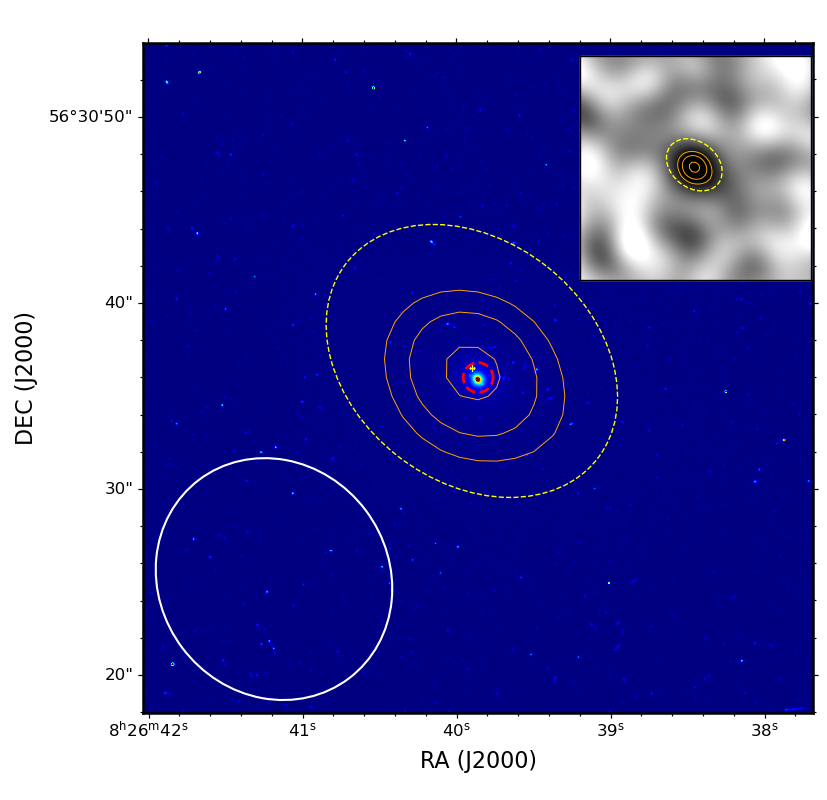}%

  \caption{ The HST image of the foreground galaxy at $z\sim0.13$. The orange contours are drawn at  $[-3,3,3.5,4] \times \sigma$ statistical levels, where $\sigma$ $\sim $ 154 $\mu$Jy beam$^{-1}$ is the RMS noise of the peak channel image. We show the corresponding peak (central) channel image in the inset.  The red dashed circle in the center is the critical curve of the lens as estimated from the SIE model of lens mass distribution. The yellow (dashed) ellipse shows the 2D fitted Gaussian to the peak channel image, with major and minor axes $17.1''\times 12.9''$ and position angle is $52^{\circ} \pm 6^{\circ}$ and the fitted center is marked by a yellow plus sign. 
  The ellipse at the bottom left is the synthesized beam of the peak channel image, with major and minor axes, $13.37''\times 12.32''$, and the position angle is $37^{\circ}$. }
\label{HST}
\end{figure*}

\section{Atomic to stellar mass ratio}

 The stellar mass of the source galaxy was not characterized due to the unavailability of the high SNR optical data \citep{Shu2017ApJ...851...48S}. In the absence of broadband photometry with high SNR, we can not estimate the stellar mass of the source galaxy in this work. However, the de-lensed magnitude of the source galaxy in HST F814W is 26.7 $\pm$ 0.4 (private communication Yiping Shu). The reference frequency of the HST F814W filter is 8100.44 $\angstrom$, which at the source redshift, $z\sim 1.3$, falls into  the SDSS u-band. We converted the magnitude of the source to u-band luminosity, $L_{\rm u} = 1.46 \times 10^{20}$ $\rm W/Hz$. \citet{Sande2015ApJ...799..125V} empirically shows that there is a strong positive correlation between dynamical mass to light ratio ($\rm log_{10}$ $M_{\rm dyn}/L_{\lambda}$) with the rest-frame color of the source.  They fit a linear function given by (see Eqn. 3 of \citealt{Sande2015ApJ...799..125V}),

\begin{equation}
    {\rm log_{10}} ~ M_{\rm dyn}/L_{\lambda} = a_{\lambda} \times {\rm C} + b_{\lambda},
    \label{eqn:light}
\end{equation}
 here C is the rest-frame (g-z) color.   We first fit a linear curve to the  (g-z) rest frame color vs redshift data taken from the top left panel of Fig.2 of \citealt{Sande2015ApJ...799..125V}. The estimated (g-z) color at $z\sim 1.3$, from the fitted linear curve, is $1.16 \pm 0.1$. The value of the coefficients,  $a_{u} = 1.89$ and $b_{u} = -1.97$,  are taken from  Table 3 of \citet{Sande2015ApJ...799..125V}. We used these fitted color and coefficient values and estimated the dynamical mass of the source galaxy using Eqn. \ref{eqn:light},  $\rm log_{10}$ $M_{\rm dyn}/M_{\odot} = 9.77 \pm 0.2 $.  \citet{Sande2015ApJ...799..125V} also compared the relation of  $M_{\rm dyn}/L$  with rest-frame color and the relation between  SED-derived stellar mass to the light ratio ($M_{*}/L$)  with the rest-frame color for a wide  range of galaxies out to $z\sim2$. They found a mean ratio of $\rm log_{10}$ $M_{*}/M_{\rm dyn} = -0.20$ with RMS scatter of 0.20 dex \citep{Sande2015ApJ...799..125V}.  This gives the stellar mass of our source galaxy of about $M_{*} = (0.38 \pm 0.11) \times 10^{10}M_{\odot}$.
 \citet{Newton2011ApJ...734..104N} have estimated the stellar mass of the  galaxies detected in SLOAN LENS ACS SURVEY (SLACS) using the broadband photometry of the and the Kroupa Initial Mass Function (IMF) \citep{Kroupa2001MNRAS.322..231K} and found that the stellar mass ($M_{*}$) of a  source galaxy (J1318-0313) at $z=1.3$,  a similar redshift as our target galaxy,  in the SLACS catalog is $M_{*}=(0.32 \pm 0.15) \times 10^{10}M_{\odot}$ (see Table 1 and 2 in \citealt{Newton2011ApJ...734..104N}), which is also nearly consistent with our findings.

  The   \hi\ to stellar mass ratio of our source galaxy is  $M_{\rm HI}/M_{*} = 2.37 \pm 0.14$, suggesting that the cold atomic gas is  higher than  the stellar component of the galaxy.  This result is in agreement with the findings of  average \hi\ to stellar mass ratio of about $1.26 \pm 0.28$ at $z\sim1$ \citep{Aditya2020Natur.586..369C} and $2.6 \pm 0.5$ at $z\sim 1.3$ \citep{Aditya2021ApJ...913L..24C} in star-forming galaxies via stacking.   However, this is in clear disagreement with the findings for the local star-forming galaxies with similar stellar mass distribution, where the average  \hi\ mass is about 35 \%  of the average stellar mass of galaxies detected in extended GALEX Arecibo SDSS Survey (xGASS) between $0.01<z<0.05$ \citep{Catinella2018MNRAS.476..875C}. There is a significant evolution of stellar mass function from $z=0$ to $z=1$, however, the  predicted HIMF shows negligible evolution in this redshift period \citep{Lagos2011MNRAS.418.1649L}. This suggests that at a given stellar mass the cold gas reservoir of galaxies will  be larger at higher redshifts.  The fact that we also found a larger \hi\ mass in comparison with the  stellar mass at redshift around 1.3 in the star-forming galaxy, suggests that an evolution of the \hi\ to stellar mass ratio from high redshift to the present epoch in star-forming galaxies.

\section{Extension of {\hi\ emission}}

  We  fit a 2D Gaussian to the  peak channel image, at $\nu = 620.0683$  MHz ($z=1.2907$), using {\sc imfit} task in {\sc casa}.  We found that the fitted major and minor axes  are $17.1'' \times 12.9''$ and the position angle (p.a) is $52^{\circ} \pm 6^{\circ}$. Fig. \ref{HST} shows the optical HST image in the F814W-band of the foreground galaxy at $z_{L}=0.13$.  The  orange lines are the $[-3,3,3.5,4] \times \sigma$ contour levels of the peak channel image ($\nu = 620.0683$ MHz), where $\sigma \sim 154 \mu$Jy beam$^{-1}$ is the RMS noise in the image. We show  the central $1.5' \times 1.5'$ region ($\sim 770$ kpc at $z\sim 1.3$) of the peak channel image in the inset. The red dashed  circle around the central foreground galaxy shows the  outer critical curve of the lens as estimated from the SIE model of lens mass distribution \citep{Shu2017ApJ...851...48S}. The measured Einstein radius was $1.01''$ and the major axis position angle  of the SIE component with respect to the North is $82^{\circ}$ \citep{Shu2017ApJ...851...48S}.  The ellipse in yellow shows the 2D fitted Gaussian to the central channel image, where the fitted center is marked by a plus sign in yellow  at $\rm RA =08^{h}26^{m}39.90 \pm 0.25^{s},\rm DEC =56^{\circ}30'36.88 \pm 1.4''$.  At the bottom left, we show the  synthesized beam, with major and minor axes are $13.37''\times12.32''$ and position angle is $37^{\circ}$. We found that the fitted source is marginally resolved along the major axis and the size along that axis, deconvolved from the synthesized beam, is $6.8''$ and the p.a of the major axis is  $52.2^{\circ} \pm 9^{\circ}$. The elongation of the \hi\ emission is tangential to the critical curve of the lens and in a different orientation compared to the orientation of the synthesized beam.  We also tried to fit a 2D Gaussian to the averaged image (Fig. \ref{fig:HI_Spectrum}) and found a similar result as that of the central channel. However, the \hi\ emission did not show any resolved structure when we fit the two nearest neighbor channels to the central channel, and the fitted size of the  major and minor axes was the same as that of the synthesized beam.

 This is probably due to the fact that the magnification of the central channel is higher than the neighbors. The reason behind this differential magnification is the position of the channel flux solid angle with respect to the caustic. If the \hi\ emission for a narrow channel overlaps with the inner lens caustic, it creates a full Einstein ring and the magnification becomes higher. However, for the neighboring channels, the \hi\ emission is either approaching or going away from the lens caustic and there is no perfect alignment happened. Due to this, the \hi\ magnification can be lower for those channels and we are unable to see the full \hi\ emission region (see  \citealt{Deane2015MNRAS.452L..49D} for a detailed discussion). We also tried with a coarser 35  km $s^{-1}$ channel resolution, but the  final SNR for that was not better than the original 24 km $s^{-1}$ channel resolution. This suggests that a narrow channel width is optimal for higher SNR detection of the \hi\ line emission, simply due to the larger magnification in some channels \citep{Deane2015MNRAS.452L..49D}.  This indicates that one needs to be careful to select an optimal channel resolution that yields a higher probability of detection. 
 
 \section{Conclusions}
\label{sec:conclusion}
 
 Strong gravitational lensing helps to study high-redshift galaxies, which can only be possible with next-generation telescopes in the absence of lensing. Here, for the first time, we report the discovery of the \hi\ 21 cm emission signal from a star-forming galaxy at $z\sim1.3$ (nearly 9 billion years ago) using uGMRT via strong gravitational lensing. This opens up a new window to probe the cold neutral gas at high redshifts. We found that the \hi\ lensing magnification is different than the optical magnification and it largely depends upon the \hi\ mass of the source galaxy. Since the \hi\ mass increases with the size of the \hi\ disk, the \hi\ magnification decreases with the \hi\ mass \citep{Blecher2019MNRAS.484.3681B}. However, the "knee" of the HIMF might shift to lower masses \citep{Lagos2011MNRAS.418.1649L} at higher redshifts and due to which the intrinsic size of the \hi\ disk (see Eqn. \ref{eq:mass_size}) is expected to be smaller, resulting into higher magnification factor. Due to this magnification boost, we would expect to detect more lensed \hi\ galaxies at high redshifts in the future.

In the absence of the high signal-to-noise ratio optical data, the stellar mass of the source galaxy can not be estimated. However, we use the relations between dynamical mass and color and the ratio between dynamical mass and stellar mass of high redshift galaxies presented in \citet{Sande2015ApJ...799..125V} and infer the stellar mass of our source galaxy. We found that the atomic-to-stellar mass ratio is significantly higher than the local star-forming galaxies. \citet{Aditya2021ApJ...913L..24C, Aditya2020Natur.586..369C} also found higher atomic-to-stellar mass ratio at high redshifts ($z \sim 1$) using stacking of many star-forming galaxies. 
This indicates that  the atomic  gas reservoir  of high redshift galaxies is large in these star-forming galaxies.

\citet{Deane2015MNRAS.452L..49D} shows that  the fraction of lensed galaxies out of all galaxies increases by 2-3 orders of magnitude from $z\sim 0.5$ to $z\sim 2$, for an integrated \hi\ flux cut at about 1.0 mJy km $s^{-1}$. The large instantaneous bandwidth of modern receivers in current and next-generation telescopes, such as uGMRT, VLA, MeerKAT, SKA1-MID; will detect a large number ($\sim 10^{4}$) of lensed \hi\ galaxies and significantly improve our understandings of the cold neutral gas reservoirs, the evolution of the HIMF and the star to gas mass ratio at high redshifts.  

{\bf Acknowledgements}
 We  thank the staff of GMRT for making this observation possible. GMRT is run by National Centre for Radio Astrophysics of the Tata Institute of Fundamental Research. We thank Yiping Shu for providing help and suggestions.   We thank the editor and the reviewers for their helpful comments and suggestions, which improve the manuscript.

\section*{Data Availability}
The data is publicly available in GMRT archive, with proposal code - 34\_066.



\bibliographystyle{mnras}
\bibliography{HI} 

\begin{thebibliography}{}
\makeatletter
\relax
\def\mn@urlcharsother{\let\do\@makeother \do\$\do\&\do\#\do\^\do\_\do\%\do\~}
\def\mn@doi{\begingroup\mn@urlcharsother \@ifnextchar [ {\mn@doi@}
  {\mn@doi@[]}}
\def\mn@doi@[#1]#2{\def\@tempa{#1}\ifx\@tempa\@empty \href
  {http://dx.doi.org/#2} {doi:#2}\else \href {http://dx.doi.org/#2} {#1}\fi
  \endgroup}
\def\mn@eprint#1#2{\mn@eprint@#1:#2::\@nil}
\def\mn@eprint@arXiv#1{\href {http://arxiv.org/abs/#1} {{\tt arXiv:#1}}}
\def\mn@eprint@dblp#1{\href {http://dblp.uni-trier.de/rec/bibtex/#1.xml}
  {dblp:#1}}
\def\mn@eprint@#1:#2:#3:#4\@nil{\def\@tempa {#1}\def\@tempb {#2}\def\@tempc
  {#3}\ifx \@tempc \@empty \let \@tempc \@tempb \let \@tempb \@tempa \fi \ifx
  \@tempb \@empty \def\@tempb {arXiv}\fi \@ifundefined
  {mn@eprint@\@tempb}{\@tempb:\@tempc}{\expandafter \expandafter \csname
  mn@eprint@\@tempb\endcsname \expandafter{\@tempc}}}

\bibitem[\protect\citeauthoryear{{Blecher}, {Deane}, {Heywood}  \&
  {Obreschkow}}{{Blecher} et~al.}{2019}]{Blecher2019MNRAS.484.3681B}
{Blecher} T.,  {Deane} R.,  {Heywood} I.,   {Obreschkow} D.,  2019, \mn@doi
  [\mnras] {10.1093/mnras/stz224}, \href
  {https://ui.adsabs.harvard.edu/abs/2019MNRAS.484.3681B} {484, 3681}

\bibitem[\protect\citeauthoryear{{Bolton}, {Burles}, {Schlegel}, {Eisenstein}
  \& {Brinkmann}}{{Bolton} et~al.}{2004}]{Bolton2004AJ....127.1860B}
{Bolton} A.~S.,  {Burles} S.,  {Schlegel} D.~J.,  {Eisenstein} D.~J.,
  {Brinkmann} J.,  2004, \mn@doi [\aj] {10.1086/382714}, \href
  {https://ui.adsabs.harvard.edu/abs/2004AJ....127.1860B} {127, 1860}

\bibitem[\protect\citeauthoryear{{Bouwens} et~al.,}{{Bouwens}
  et~al.}{2009}]{Bouwens2009ApJ...705..936B}
{Bouwens} R.~J.,  et~al., 2009, \mn@doi [\apj] {10.1088/0004-637X/705/1/936},
  \href {https://ui.adsabs.harvard.edu/abs/2009ApJ...705..936B} {705, 936}

\bibitem[\protect\citeauthoryear{{Briggs}}{{Briggs}}{1995}]{Briggs1995AAS...18711202B}
{Briggs} D.~S.,  1995, in American Astronomical Society Meeting Abstracts. p.
  112.02

\bibitem[\protect\citeauthoryear{{Brown} \& {Vanden Bout}}{{Brown} \& {Vanden
  Bout}}{1991}]{Brown1991AJ....102.1956B}
{Brown} R.~L.,  {Vanden Bout} P.~A.,  1991, \mn@doi [\aj] {10.1086/116017},
  \href {https://ui.adsabs.harvard.edu/abs/1991AJ....102.1956B} {102, 1956}

\bibitem[\protect\citeauthoryear{{CHIME Collaboration} et~al.,}{{CHIME
  Collaboration} et~al.}{2022}]{CHIME2022arXiv220201242C}
{CHIME Collaboration} et~al., 2022, arXiv e-prints, \href
  {https://ui.adsabs.harvard.edu/abs/2022arXiv220201242C} {p. arXiv:2202.01242}

\bibitem[\protect\citeauthoryear{{Catinella} et~al.,}{{Catinella}
  et~al.}{2018}]{Catinella2018MNRAS.476..875C}
{Catinella} B.,  et~al., 2018, \mn@doi [\mnras] {10.1093/mnras/sty089}, \href
  {https://ui.adsabs.harvard.edu/abs/2018MNRAS.476..875C} {476, 875}

\bibitem[\protect\citeauthoryear{{Chengalur}, {Braun}  \&
  {Wieringa}}{{Chengalur} et~al.}{2001}]{Chengalur2001A&A...372..768C}
{Chengalur} J.~N.,  {Braun} R.,   {Wieringa} M.,  2001, \mn@doi [\aap]
  {10.1051/0004-6361:20010547}, \href
  {https://ui.adsabs.harvard.edu/abs/2001A&A...372..768C} {372, 768}

\bibitem[\protect\citeauthoryear{{Chowdhury}, {Kanekar}, {Chengalur}, {Sethi}
  \& {Dwarakanath}}{{Chowdhury} et~al.}{2020}]{Aditya2020Natur.586..369C}
{Chowdhury} A.,  {Kanekar} N.,  {Chengalur} J.~N.,  {Sethi} S.,   {Dwarakanath}
  K.~S.,  2020, \mn@doi [\nat] {10.1038/s41586-020-2794-7}, \href
  {https://ui.adsabs.harvard.edu/abs/2020Natur.586..369C} {586, 369}

\bibitem[\protect\citeauthoryear{{Chowdhury}, {Kanekar}, {Das}, {Dwarakanath}
  \& {Sethi}}{{Chowdhury} et~al.}{2021}]{Aditya2021ApJ...913L..24C}
{Chowdhury} A.,  {Kanekar} N.,  {Das} B.,  {Dwarakanath} K.~S.,   {Sethi} S.,
  2021, \mn@doi [\apjl] {10.3847/2041-8213/abfcc7}, \href
  {https://ui.adsabs.harvard.edu/abs/2021ApJ...913L..24C} {913, L24}

\bibitem[\protect\citeauthoryear{{Cornwell}, {Golap}  \&
  {Bhatnagar}}{{Cornwell} et~al.}{2008}]{Cornwell2008ISTSP...2..647C}
{Cornwell} T.~J.,  {Golap} K.,   {Bhatnagar} S.,  2008, \mn@doi [IEEE Journal
  of Selected Topics in Signal Processing] {10.1109/JSTSP.2008.2005290}, \href
  {https://ui.adsabs.harvard.edu/abs/2008ISTSP...2..647C} {2, 647}

\bibitem[\protect\citeauthoryear{{Cowie}, {Songaila}, {Hu}  \& {Cohen}}{{Cowie}
  et~al.}{1996}]{Cowie1996AJ....112..839C}
{Cowie} L.~L.,  {Songaila} A.,  {Hu} E.~M.,   {Cohen} J.~G.,  1996, \mn@doi
  [\aj] {10.1086/118058}, \href
  {https://ui.adsabs.harvard.edu/abs/1996AJ....112..839C} {112, 839}

\bibitem[\protect\citeauthoryear{{Deane}, {Obreschkow}  \& {Heywood}}{{Deane}
  et~al.}{2015}]{Deane2015MNRAS.452L..49D}
{Deane} R.~P.,  {Obreschkow} D.,   {Heywood} I.,  2015, \mn@doi [\mnras]
  {10.1093/mnrasl/slv086}, \href
  {https://ui.adsabs.harvard.edu/abs/2015MNRAS.452L..49D} {452, L49}

\bibitem[\protect\citeauthoryear{{Fern{\'a}ndez} et~al.,}{{Fern{\'a}ndez}
  et~al.}{2016}]{Fernandez2016ApJ...824L...1F}
{Fern{\'a}ndez} X.,  et~al., 2016, \mn@doi [\apjl]
  {10.3847/2041-8205/824/1/L1}, \href
  {https://ui.adsabs.harvard.edu/abs/2016ApJ...824L...1F} {824, L1}

\bibitem[\protect\citeauthoryear{{Hopkins} \& {Beacom}}{{Hopkins} \&
  {Beacom}}{2006}]{Hopkins2006ApJ...651..142H}
{Hopkins} A.~M.,  {Beacom} J.~F.,  2006, \mn@doi [\apj] {10.1086/506610}, \href
  {https://ui.adsabs.harvard.edu/abs/2006ApJ...651..142H} {651, 142}

\bibitem[\protect\citeauthoryear{{Hunt}, {Pisano}  \& {Edel}}{{Hunt}
  et~al.}{2016}]{Hunt2016AJ....152...30H}
{Hunt} L.~R.,  {Pisano} D.~J.,   {Edel} S.,  2016, \mn@doi [\aj]
  {10.3847/0004-6256/152/2/30}, \href
  {https://ui.adsabs.harvard.edu/abs/2016AJ....152...30H} {152, 30}

\bibitem[\protect\citeauthoryear{{Jones}, {Haynes}, {Giovanelli}  \&
  {Moorman}}{{Jones} et~al.}{2018}]{Jones2018MNRAS.477....2J}
{Jones} M.~G.,  {Haynes} M.~P.,  {Giovanelli} R.,   {Moorman} C.,  2018,
  \mn@doi [\mnras] {10.1093/mnras/sty521}, \href
  {https://ui.adsabs.harvard.edu/abs/2018MNRAS.477....2J} {477, 2}

\bibitem[\protect\citeauthoryear{{Kormann}, {Schneider}  \&
  {Bartelmann}}{{Kormann} et~al.}{1994}]{Kormann1994A&A...284..285K}
{Kormann} R.,  {Schneider} P.,   {Bartelmann} M.,  1994, \aap, \href
  {https://ui.adsabs.harvard.edu/abs/1994A&A...284..285K} {284, 285}

\bibitem[\protect\citeauthoryear{{Kroupa}}{{Kroupa}}{2001}]{Kroupa2001MNRAS.322..231K}
{Kroupa} P.,  2001, \mn@doi [\mnras] {10.1046/j.1365-8711.2001.04022.x}, \href
  {https://ui.adsabs.harvard.edu/abs/2001MNRAS.322..231K} {322, 231}

\bibitem[\protect\citeauthoryear{{Lagos}, {Baugh}, {Lacey}, {Benson}, {Kim}  \&
  {Power}}{{Lagos} et~al.}{2011}]{Lagos2011MNRAS.418.1649L}
{Lagos} C. D.~P.,  {Baugh} C.~M.,  {Lacey} C.~G.,  {Benson} A.~J.,  {Kim}
  H.-S.,   {Power} C.,  2011, \mn@doi [\mnras]
  {10.1111/j.1365-2966.2011.19583.x}, \href
  {https://ui.adsabs.harvard.edu/abs/2011MNRAS.418.1649L} {418, 1649}

\bibitem[\protect\citeauthoryear{{Le Floc'h} et~al.,}{{Le Floc'h}
  et~al.}{2005}]{LeFloch2005ApJ...632..169L}
{Le Floc'h} E.,  et~al., 2005, \mn@doi [\apj] {10.1086/432789}, \href
  {https://ui.adsabs.harvard.edu/abs/2005ApJ...632..169L} {632, 169}

\bibitem[\protect\citeauthoryear{{Madau} \& {Dickinson}}{{Madau} \&
  {Dickinson}}{2014}]{Madau2014ARA&A..52..415M}
{Madau} P.,  {Dickinson} M.,  2014, \mn@doi [\araa]
  {10.1146/annurev-astro-081811-125615}, \href
  {https://ui.adsabs.harvard.edu/abs/2014ARA&A..52..415M} {52, 415}

\bibitem[\protect\citeauthoryear{{Maddox}, {Hess}, {Obreschkow}, {Jarvis}  \&
  {Blyth}}{{Maddox} et~al.}{2015}]{Maddox2015MNRAS.447.1610M}
{Maddox} N.,  {Hess} K.~M.,  {Obreschkow} D.,  {Jarvis} M.~J.,   {Blyth} S.~L.,
   2015, \mn@doi [\mnras] {10.1093/mnras/stu2532}, \href
  {https://ui.adsabs.harvard.edu/abs/2015MNRAS.447.1610M} {447, 1610}

\bibitem[\protect\citeauthoryear{{Newton}, {Marshall}, {Treu}, {Auger},
  {Gavazzi}, {Bolton}, {Koopmans}  \& {Moustakas}}{{Newton}
  et~al.}{2011}]{Newton2011ApJ...734..104N}
{Newton} E.~R.,  {Marshall} P.~J.,  {Treu} T.,  {Auger} M.~W.,  {Gavazzi} R.,
  {Bolton} A.~S.,  {Koopmans} L. V.~E.,   {Moustakas} L.~A.,  2011, \mn@doi
  [\apj] {10.1088/0004-637X/734/2/104}, \href
  {https://ui.adsabs.harvard.edu/abs/2011ApJ...734..104N} {734, 104}

\bibitem[\protect\citeauthoryear{{Obreschkow}, {Croton}, {De Lucia}, {Khochfar}
   \& {Rawlings}}{{Obreschkow} et~al.}{2009}]{Obreschkow2009ApJ...698.1467O}
{Obreschkow} D.,  {Croton} D.,  {De Lucia} G.,  {Khochfar} S.,   {Rawlings} S.,
   2009, \mn@doi [\apj] {10.1088/0004-637X/698/2/1467}, \href
  {https://ui.adsabs.harvard.edu/abs/2009ApJ...698.1467O} {698, 1467}

\bibitem[\protect\citeauthoryear{{Offringa} \& {Smirnov}}{{Offringa} \&
  {Smirnov}}{2017}]{Offringa2017MNRAS.471..301O}
{Offringa} A.~R.,  {Smirnov} O.,  2017, \mn@doi [\mnras]
  {10.1093/mnras/stx1547}, \href
  {https://ui.adsabs.harvard.edu/abs/2017MNRAS.471..301O} {471, 301}

\bibitem[\protect\citeauthoryear{{Offringa}, {van de Gronde}  \&
  {Roerdink}}{{Offringa} et~al.}{2012}]{Offringa2012A&A...539A..95O}
{Offringa} A.~R.,  {van de Gronde} J.~J.,   {Roerdink} J.~B.~T.~M.,  2012,
  \mn@doi [\aap] {10.1051/0004-6361/201118497}, \href
  {https://ui.adsabs.harvard.edu/abs/2012A&A...539A..95O} {539, A95}

\bibitem[\protect\citeauthoryear{{Offringa} et~al.,}{{Offringa}
  et~al.}{2014}]{Offringa2014MNRAS.444..606O}
{Offringa} A.~R.,  et~al., 2014, \mn@doi [\mnras] {10.1093/mnras/stu1368},
  \href {https://ui.adsabs.harvard.edu/abs/2014MNRAS.444..606O} {444, 606}

\bibitem[\protect\citeauthoryear{{Oguri}}{{Oguri}}{2010}]{Oguri2010PASJ...62.1017O}
{Oguri} M.,  2010, \mn@doi [\pasj] {10.1093/pasj/62.4.1017}, \href
  {https://ui.adsabs.harvard.edu/abs/2010PASJ...62.1017O} {62, 1017}

\bibitem[\protect\citeauthoryear{{Perley} \& {Butler}}{{Perley} \&
  {Butler}}{2017}]{Perley2017ApJS..230....7P}
{Perley} R.~A.,  {Butler} B.~J.,  2017, \mn@doi [\apjs]
  {10.3847/1538-4365/aa6df9}, \href
  {https://ui.adsabs.harvard.edu/abs/2017ApJS..230....7P} {230, 7}

\bibitem[\protect\citeauthoryear{{Planck Collaboration} et~al.,}{{Planck
  Collaboration} et~al.}{2016}]{Planck2016A&A...594A..13P}
{Planck Collaboration} et~al., 2016, \mn@doi [\aap]
  {10.1051/0004-6361/201525830}, \href
  {https://ui.adsabs.harvard.edu/abs/2016A&A...594A..13P} {594, A13}

\bibitem[\protect\citeauthoryear{{Shu} et~al.,}{{Shu}
  et~al.}{2015}]{Shu2015ApJ...803...71S}
{Shu} Y.,  et~al., 2015, \mn@doi [\apj] {10.1088/0004-637X/803/2/71}, \href
  {https://ui.adsabs.harvard.edu/abs/2015ApJ...803...71S} {803, 71}

\bibitem[\protect\citeauthoryear{{Shu} et~al.,}{{Shu}
  et~al.}{2017}]{Shu2017ApJ...851...48S}
{Shu} Y.,  et~al., 2017, \mn@doi [\apj] {10.3847/1538-4357/aa9794}, \href
  {https://ui.adsabs.harvard.edu/abs/2017ApJ...851...48S} {851, 48}

\bibitem[\protect\citeauthoryear{{Vieira} et~al.,}{{Vieira}
  et~al.}{2013}]{Vieira2013Natur.495..344V}
{Vieira} J.~D.,  et~al., 2013, \mn@doi [\nat] {10.1038/nature12001}, \href
  {https://ui.adsabs.harvard.edu/abs/2013Natur.495..344V} {495, 344}

\bibitem[\protect\citeauthoryear{{Walter} et~al.,}{{Walter}
  et~al.}{2020}]{Walter2020ApJ...902..111W}
{Walter} F.,  et~al., 2020, \mn@doi [\apj] {10.3847/1538-4357/abb82e}, \href
  {https://ui.adsabs.harvard.edu/abs/2020ApJ...902..111W} {902, 111}

\bibitem[\protect\citeauthoryear{{Wang}, {Koribalski}, {Serra}, {van der
  Hulst}, {Roychowdhury}, {Kamphuis}  \& {Chengalur}}{{Wang}
  et~al.}{2016}]{Wang2016MNRAS.460.2143W}
{Wang} J.,  {Koribalski} B.~S.,  {Serra} P.,  {van der Hulst} T.,
  {Roychowdhury} S.,  {Kamphuis} P.,   {Chengalur} J.~N.,  2016, \mn@doi
  [\mnras] {10.1093/mnras/stw1099}, \href
  {https://ui.adsabs.harvard.edu/abs/2016MNRAS.460.2143W} {460, 2143}

\bibitem[\protect\citeauthoryear{{Wieringa}, {de Bruyn}  \&
  {Katgert}}{{Wieringa} et~al.}{1992}]{Wieringa1992A&A...256..331W}
{Wieringa} M.~H.,  {de Bruyn} A.~G.,   {Katgert} P.,  1992, \aap, \href
  {https://ui.adsabs.harvard.edu/abs/1992A&A...256..331W} {256, 331}

\bibitem[\protect\citeauthoryear{{Zwaan}}{{Zwaan}}{2000}]{Zwaan2000PhDT..........Z}
{Zwaan} M.~A.,  2000, PhD thesis, University of Groningen, Netherlands

\bibitem[\protect\citeauthoryear{{van de Sande}, {Kriek}, {Franx}, {Bezanson}
  \& {van Dokkum}}{{van de Sande} et~al.}{2015}]{Sande2015ApJ...799..125V}
{van de Sande} J.,  {Kriek} M.,  {Franx} M.,  {Bezanson} R.,   {van Dokkum}
  P.~G.,  2015, \mn@doi [\apj] {10.1088/0004-637X/799/2/125}, \href
  {https://ui.adsabs.harvard.edu/abs/2015ApJ...799..125V} {799, 125}

\makeatother
\end{thebibliography}







\bsp	
\label{lastpage}
\end{document}